\documentclass[twocolumn]{svjour3}          
\smartqed  
\usepackage{graphicx}
\usepackage{multirow}
\usepackage{pifont}
\usepackage{listings}
\usepackage{url}

\newcommand{\cmark}{\ding{51}}%
\newcommand{\xmark}{\ding{55}}%

\graphicspath{ {images/} }

\newcommand\YAMLcolonstyle{\color{red}\mdseries}
\newcommand\YAMLkeystyle{\color{black}\bfseries}
\newcommand\YAMLvaluestyle{\color{blue}\mdseries}

\makeatletter

\newcommand\language@yaml{yaml}

\expandafter\expandafter\expandafter\lstdefinelanguage
\expandafter{\language@yaml}
{
  keywords={true,false,null,y,n},
  keywordstyle=\color{darkgray}\bfseries,
  basicstyle=\YAMLkeystyle,                                 
  sensitive=false,
  comment=[l]{\#},
  morecomment=[s]{/*}{*/},
  commentstyle=\color{purple}\ttfamily,
  stringstyle=\YAMLvaluestyle\ttfamily,
  moredelim=[l][\color{orange}]{\&},
  moredelim=[l][\color{magenta}]{*},
  moredelim=**[il][\YAMLcolonstyle{:}\YAMLvaluestyle]{:},   
  morestring=[b]',
  morestring=[b]",
  literate =    {---}{{\ProcessThreeDashes}}3
                {>}{{\textcolor{red}\textgreater}}1     
                {|}{{\textcolor{red}\textbar}}1 
                {\ -\ }{{\mdseries\ -\ }}3,
}

\lst@AddToHook{EveryLine}{\ifx\lst@language\language@yaml\YAMLkeystyle\fi}
\makeatother

\newcommand\ProcessThreeDashes{\llap{\color{cyan}\mdseries-{-}-}}

\usepackage{color}

\lstnewenvironment{python}[1][]
{
\pythonstyle
\lstset{#1}
}
{}

\newcommand\pythoninline[1]{{\pythonstyle\lstinline!#1!}}

\begin{document}

\title{\texttt{umd-verification}: Automation of Software Validation for the EGI federated e-Infrastructure
}


\author{Pablo Orviz Fern\'andez \and
        Jo\~ao Pina \and \'Alvaro L\'opez Garc\'ia \and Isabel Campos Plasencia \and M\'ario David \and Jorge Gomes
}


\institute{Pablo Orviz Fern\'andez - \'Alvaro L\'opez Garc\'ia - Isabel Campos Plasencia \at
              Instituto de F\'isica de Cantabria (CSIC), Santander, Spain \\
              \email{orviz@ifca.unican.es}           
           \and
           Jo\~ao Pina - M\'ario David - Jorge Gomes \at
              Laborat\'orio de Instrumenta\c c\~ao e F\'isica Experimental de Part\'iculas (LIP), Lisboa, Portugal
}


\date{\textbf{This is the author’s pre-print version of this work. The final publication is available at \url{http://dx.doi.org/10.1007/s10723-018-9454-2}}}

\maketitle

\begin{abstract}
\begin{sloppypar}
Supporting e-Science in the EGI e-Infrastructure requires extensive and reliable
software, for advanced computing use, deployed across over approximately 300
European and worldwide data centers. The Unified Middleware Distribution (UMD)
and Cloud Middleware Distribution (CMD) are the channels to deliver the software
for the EGI e-Infrastructure consumption. The software is compiled, validated and
distributed following the Software Provisioning Process (SWPP), where the Quality
Criteria (QC) definition sets the minimum quality requirements for EGI acceptance.
The growing number of software components currently existing within UMD and CMD 
distributions hinders the application of the traditional, manual-based validation 
mechanisms, thus driving the adoption of automated solutions. This paper presents 
\texttt{umd-verification}, an open-source tool that enforces the fulfillment of the
QC requirements in an automated way for the continuous validation of the software
products for scientific disposal. The \texttt{umd-verification} tool has been
successfully integrated within the SWPP pipeline and is progressively supporting 
the full validation of the products in the UMD and CMD repositories. While the cost
of supporting new products is dependant on the availability of Infrastructure as
Code solutions to take over the deployment and high test coverage, the results
obtained for the already integrated products are promising, as the time invested in
the validation of products has been drastically reduced. Furthermore, automation
adoption has brought along benefits for the reliability of the process, such as the
removal of human-associated errors or the risk of regression of previously tested 
functionalities.
\end{sloppypar}
\keywords{Automation \and Software Verification and Validation \and Software Quality
Assurance \and Software Quality Control \and Software Testing \and 
Continuous Integration}
\end{abstract}

%
%
\section{Introduction}
\label{intro}
\begin{sloppypar}
EGI \cite{egi-web} federates computing and data resources, mainly hosted in Europe,
to satisfy common and specific research requirements gathered from multidisciplinary
scientific communities. EGI operates as an e-Infrastructure \cite{Andronico2011} that
exploits complex data-intensive Grid and Cloud computing services \cite{Shamsi2013,Montes2012}
through the Unified Middleware Distribution
(UMD)\footnote{http://repository.egi.eu/sw/production/umd/} and the Cloud
Middleware Distribution (CMD)\footnote{http://repository.egi.eu/sw/production/cmd-os/}\footnote{http://repository.egi.eu/sw/production/cmd-one/} official releases,
respectively.
\end{sloppypar}
\begin{figure*}[t]
\centerline{\includegraphics[scale=0.5]{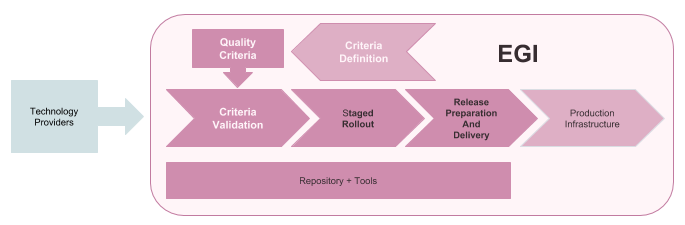}}
\caption{EGI Software Provisioning Process (SWPP).
\label{fig_swpp}}
\end{figure*}

The UMD and CMD distributions provide repositories to distribute software in the form of
Linux packages that are provisioned by external \textit{technology providers} (TPs). 
The environments in which the software has been developed are not under the control nor
monitored by EGI, thus there is no guarantee that the software is reliable enough for
the production infrastructures. Therefore, EGI invests on a validation effort
for the incoming software products to lessen the odds of disruption. The Software 
Provisioning Process (SWPP) \cite{Mario2014}, schematized in Figure \ref{fig_swpp},
guides the EGI software delivery through UMD and CMD distributions, encompassing the 
i) \textit{validation of the conformance criteria}, scope of the present paper, the ii)
\textit{staged rollout} phase, which takes over the deployment and user-level testing
on production facilities, and, finally, the iii) \textit{release to production}, resulting
in the software release preparation and delivery.

During the validation of the conformance criteria phase, every piece of software is
deployed and tested to detect any malfunction or deviation from the design specification.
The procedure of validation is governed by the Quality Criteria (QC) definition,
which enforces the quality requirements that any software released under UMD and CMD
distributions must comply. The validation phase appears as the most time-consuming task
within the SWPP since a major effort is spent on dealing with the deployment
peculiarities of each software component, as well as in ensuring a minimal testing coverage.
Consequently, the validation process requires some modernization that optimizes the effort
invested, being able to respond accurately to the growing needs of UMD and CMD consumers.


The remainder of this paper outlines the automated solution implemented to speed up the
process of conformance criteria validation for UMD and CMD products. Section 
\ref{sec:automation} introduces the difficulties of preserving the traditional validation 
process, presenting automation as a suitable solution for the EGI QC enforcement. Section 
\ref{section:related-work} contextualizes the QC validation in the software engineering
literature, emphasizing the role of automation in the methodologies reviewed. Section
\ref{section:implementation} introduces the new tool, \texttt{umd-verification}, that
drives the QC validation process in an automated fashion. Finally, Section
\ref{section:evidence} highlights the proven advancements obtained after applying the
\texttt{umd-verification} tool in the EGI QC validation process.

\section{Boosting the validation process}
\label{sec:automation}

\subsection{Statement of the problem}
\label{sec:automation:problem}

\begin{figure*}[t]
\centerline{\includegraphics[scale=0.5]{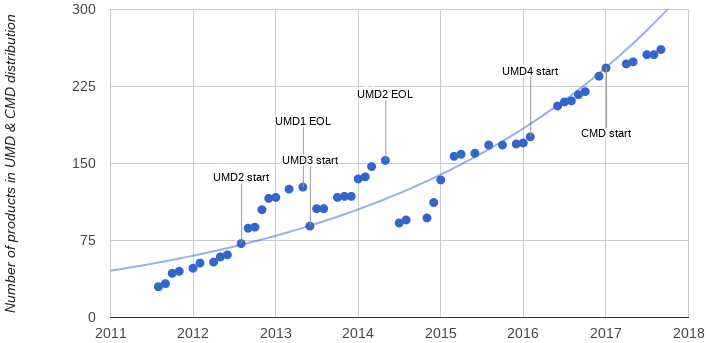}}
\caption{Trend graph showing the number of products supported in the EGI production
repositories (UMD and CMD). The incremental trend is interrupted by the end-of-life (EOL)
cycles, which are rapidly recovered as a result of the parallel start of the subsequent
major release version. At this point in time, the incoming UMD major release progressively
adopts, following the validation process, the products previously existing.}
\label{fig_umd_supported_products_trendline}
\end{figure*}

An analysis of the evolution of EGI software product catalogue, outlined in Figure
\ref{fig_umd_supported_products_trendline}, shows a \textbf{growing trend in the number
of products being supported} since the first release of the UMD distribution, UMD-1. The
underlying reasons behind this growth are mainly the evolving technology demands coming
from the scientific communities leveraging the EGI e-Infrastructure. Recently, these user 
requirements resulted in the release of the CMD distribution --as Cloud computing became
a popular technology for research computation--, thus considerably increasing the number of
products supported in the EGI production catalogue.

It is important to underline that Figure \ref{fig_umd_supported_products_trendline}
only shows the total products, not the actual validations being performed. The number of 
validations averagely increases in a factor of 2, depending on the number of operating system
(OS) distributions supported within each UMD major release. Table \ref{table:oses} shows the 
specific OSes supported throughout the UMD and CMD major releases, which in some cases raised
up to 3 different OS distributions.

Addressing the growing needs with the former validation process resulted in delays within the
SWPP chain, leading to extreme situations where a product release was disregarded and
superseded by a subsequent release while queued at this stage. According to \cite{Mario2014},
the validation of the conformance criteria phase was driven by a team of 15-20 testers, 
each taking over the product validation process based on their expertise. The process was fully
manual, with a typical estimated time completion of 1 or 2 working days for each software
validation. Thereby, the traditional approach of QC validation is only sustainable as long as
the \texttt{manpower:number of products} ratio remains balanced, which is likely to become
unsustainable over time, based on the trend discussed above.

\begin{table}
\begin{tabular}{|c|c|c|}
\hline
Distribution & Major release & OSes\\
\hline
\hline
\multirow{5}{*}{UMD} & UMD-1 & Scientific Linux 5\\
\cline{2-3}
& \multirow{2}{*}{UMD-2} & Scientific Linux 5\\
& {} & Debian Squeeze\\
\cline{2-3}
& \multirow{3}{*}{UMD-3} & Scientific Linux 6\\
& {} & Scientific Linux 5*\\
& {} & Debian Squeeze*\\
\cline{2-3}
& \multirow{2}{*}{UMD-4} & CentOS7\\
& {} & Scientific Linux 6\\
\hline
\multirow{2}{*}{CMD} & \multirow{2}{*}{CMD-OS} & CentOS7\\
& {} & Ubuntu 16.04\\
& {} & Ubuntu 14.04*\\
\cline{2-3}
& CMD-ONE & CentOS7\\
\hline
\end{tabular}
\caption{Operating systems (OSes) supported throughout UMD and CMD distributions lifetime. The
support for the OSes marked with an '*' were dropped during the associated release.}
\label{table:oses}
\end{table}

\subsection{Embracing automation}
\label{sec:automation:embrace-auto}

The adoption of automation seems to be an obvious choice to address the delays within
the validation phase. This statement rests on the following assumptions.

\subsubsection*{Manpower}
Taking into consideration the above-mentioned fact of requiring 2 working days for each
\texttt{product:OS} validation, in the likely event of having 20 queued products supported
in 2 different OSes, approximately 80 working days would be needed to complete their
validation. Distributing the load among the 15-20 testers, the process would take roughly
a full-time week of work for all the members in the validation team.

While manual validation strongly relies on manpower, an automated approach would only
require effort when supporting new products, as the maintenance costs are not highly
demanding. Following the current trend, the manual process will soon not scale, requiring
more and more testers to satisfy the incoming rate of products.

\subsubsection*{Expert dependence}
The good progress of a manual software validation is driven by seasoned teams, usually
system administrators from resource centers taking part in the EGI e-Infrastructure that
are highly familiarized with the evaluated product. For this type of validation, 
technically skilled experts are required to work around unpredictable issues not addressed
in the documentation provided by the TPs.

The programmatic implementation of a product validation would only require from expert
knowledge the first time it is set up. Once in place, the process could be taken over by
non-expert testers since most of the complexity is hidden. This represents a much lighter
dependence on skilled testers, being better positioned towards risk of knowledge loss.

\subsubsection*{Reproducibility and repeatability: fighting the human factor}
In the context of mechanical or repetitive processes, the likelihood of human error is
substantially higher than when the same process is performed in an automated environment.
Whilst automated processes are predictable, humans are not able to work
with the same level of consistency.

The deterministic nature of computational solutions makes easier to achieve a high level of
repeatability in the results obtained when applied to the same input data. Conversely, the
same task performed manually could lead to unexpected outcomes as the procedure may not be
strictly fulfilled in consecutive iterations. Moreover, the programmatic implementation of
a clearly defined iterative procedure, such as EGI's QC, makes the solution reproducible. 
Therefore, subsequent executions shall obtain the same results as long as the validation
process is taken over under the same conditions, regardless of the tester.

\subsubsection*{Time efficiency}
Automation streamlines the time required to complete a task. Time efficiency is usually 
associated with automation since it allows to meet strict deadlines or even increase the
number of tests that could be performed in the same time slot, resulting in higher test
coverages.

\begin{table*}[htbp]
\begin{tabular*}{\textwidth}{|c@{\extracolsep\fill}cccc|}
\hline
Category & ID & Check & Critical & Automated \\
\hline
\hline
\multirow{ 2}{*}{Documentation} & QC\_DOC\_1 & Release notes provisioning & \cmark & \xmark \\
& QC\_DOC\_2 & User documentation & \cmark & \xmark\\
& QC\_DOC\_3 & API documentation & \xmark & \xmark\\
& QC\_DOC\_4 & Admin documentation & \cmark & \xmark\\
& QC\_DOC\_5 & Software license & \cmark & \cmark\\
\hline
\multirow{ 2}{*}{Installation} & QC\_DIST\_1 & Binary distribution (RPM, DEB) & \cmark & \cmark \\
& QC\_UPGRADE\_1 & Upgrade previous working version & \xmark & \cmark \\
\hline
\multirow{ 2}{*}{Security} & QC\_SEC\_1 & X.509 certificate support & \cmark & \cmark \\
& QC\_SEC\_2 & SHA-2 certificate support & \cmark & \cmark \\
& QC\_SEC\_3 & RFC proxy support & \xmark & \cmark \\
& QC\_SEC\_4 & ARGUS auth integration & \xmark & \cmark \\
& QC\_SEC\_5 & World writable files & \cmark & \cmark \\
& QC\_SEC\_6 & Passwords in world readable files & \cmark & \cmark \\
\hline
\multirow{ 2}{*}{Information Model} & QC\_INFO\_1 & GLUE schema 1.3 support & \xmark & \cmark \\
& QC\_INFO\_2 & GLUE schema 2.0 support  & \cmark & \cmark \\
& QC\_INFO\_3 & Middleware version & \xmark & \cmark \\
\hline
\multirow{ 2}{*}{Operations} & QC\_MON\_1 & Service probes & \xmark & \cmark \\
& QC\_ACC\_1 & Accounting records & \cmark & \cmark \\
\hline
\multirow{ 1}{*}{Support} & QC\_SUPPORT\_1 & Bug tracking system & \cmark & \cmark \\
\hline
\multirow{ 2}{*}{Specific QC} & QC\_FUNC\_1 & Basic functionality test & \cmark & \cmark \\
& QC\_FUNC\_2 & New feature/bug fixes test & \xmark & \cmark \\
\hline
\end{tabular*}
\caption{Quality Criteria (QC) requirements.}
\label{table:qc}
\end{table*}

\subsection{Automation assessment of the EGI Quality Criteria requirements}
\label{sec:automation:qc}

Early introduced, the Quality Criteria (QC) document drives the validation of software products
within the SWPP workflow. It defines the quality requirements that a given product has to
fulfill in order to be considered ready for the subsequent \textit{staged rollout} phase.
The document is continuously evolving and it is currently on the 7th release \cite{egi-qc-web}.

Table \ref{table:qc} lists the quality requirements, their associated criticality and the
possibilities of automation. Requirements cover the minimum criteria for EGI acceptance, 
and are grouped in seven broad categories: i) \textit{documentation}, ii) 
\textit{installation}, covering the full deployment of the product, iii) \textit{security},
iv) \textit{information model}, which validates the outbound data published by the
information service, v) \textit{operations}, which groups probes related to EGI 
e-Infrastructure, vi) \textit{support} channels and vii) \textit{other specific criteria}, 
useful to extend the functionality and integration testing coverage. 

As depicted in the table, the only requirements that need human interaction are the ones
related to the analysis of the documentation (\texttt{QC\_DOC\_x}): one could address
programmatically the existence of the required documentation but not the suitability of
its content. Nevertheless, the \texttt{QC\_DOC\_x} requirements seldom involve major
changes --only when products are included for the first time--, commonly appearing
as minimal improvements when it comes to software updates.

Once the requirements suitable for automation are identified and defined, the process to 
tackle them has to be implemented. From the requirement list, deployment and testing 
related tasks are the most complex and as such will be thoroughly covered in the next
sections.

%
%
\section{Related work}
\label{section:related-work}

Free and open-source software operating systems, such as Linux distributions, rely
on packages to distribute the software. Packages are archives containing the binaries,
configuration files and dependency information, accessible through online repositories. 
Software packages can be found in different formats attached to a specific Linux distribution, 
although there are recent solutions that containeirise software applications, bundling their
dependencies, to make them installable across all major Linux 
distributions \cite{AppImage,snap,flatpak}. Most quality-aware distributions have quality
control policies for package creation \cite{debian-policy} and dependency 
resolution \cite{debian-piuparts}. Likewise Linux operating systems, the software distributed
through UMD and CMD releases are in the form of packages, which also are passed through a quality
control process. As the latter are lighter distributions, they can afford to go a step further in
the software validation, imposing deployment and testing requirements.

Software validation is the process that checks that the software satisfies its intended
use, in conformance with the requirements coming from the end users. Tightly related and 
complemented by the software verification process, they together address \textit{"all
software life cycle processes including acquisition, supply, development, operation and
maintenance"}, as defined in the IEEE Standard for Software Verification and Validation
(V\&V) \cite{IEEE-VV-standard}. V\&V are commonplace concepts in software engineering 
literature, but these terms are often used interchangeably in practice \cite{ryan2017use}.
Indeed, both processes serve different purposes since verification is linked to the early
stages of the software development life cycle, focusing on building the software
correctly, while validation is commonly placed at the end of the development process,
providing \textit{"evidence that the software and its associated products satisfy system
requirements allocated to software at the end of each life cycle, solve the right 
problem, and satisfy intended use and user needs"}. The V\&V distinction is consistent
with major systems engineering processes for software development, such as the Capability
Maturity Model Integrated 
\cite{CMMI-development-2010,CMMI-services-2010,CMMI-acquisition-2010},
organized in maturity levels, where software V\&V practices are addressed at the higher
levels of the process \cite{Monteiro2009}. 

A practical way to put V\&V into action is referring to the type of testing associated
to each process. Software verification implies the static analysis of the source code, 
requirements and design documents for defect detection via inspections, walkthroughs and
reviews \cite{german2003software}. Conversely, software validation requires the software
to be in operation mode to be tested, so it is identified with the dynamic behaviour of
the source code. There are different test-case design methodologies to tackle the dynamic
analysis of a software component but all fall in the category of so-called 
\textit{black-box testing}. In this type of testing the test-cases are data or 
input/output driven, as the internal structure of the software is not of interest at this
stage. In this regard, Myers et al. \cite{Myers2012} group under the term
\textit{higher-order testing} the type of black-box testing methods --function, system,
installation, integration, acceptance-- that aim to detect defects, from the user's
perspective, by categorizing the test cases in which the software shall be exposed. The
outcome is a quality criteria that guide the software validation.

The ultimate goal of software validation is to increase the reliability of the systems
being delivered to the users. Nevertheless, in software validation, the economics of testing 
shall be carefully considered. On the one hand, inadequate investment may imply solving 
defects at later stages. Quoting from Perry's book \cite{perry2007effective}, \textit{"it is
at least 10 times as costly to correct an error after coding as before, and 100 times as
costly to correct a production error"}. On the other hand, a generous effort may lead to 
increased project costs \cite{kit1995software}, not estimated in the project design, and 
delays in the release dates \cite{huang2005optimal}. Therefore, measuring the 
cost-effectiveness of the testing process does not only imply stopping at the optimum point
where the cost of testing does not exceed the value obtained from the defects uncovered, but
also focusing on the valuable features first within the appropriate testing phase in the life
cycle \cite{bullock2000calculating}.

Test automation is gaining momentum as a way to decrease the costs and time associated to
software testing tasks. Process efficiency gets improved as automation optimizes the
execution time of testing, maximizing the test coverage as more testing could be performed
in less time \cite{saglietti2010automated}. The augmentation of the test coverage strengthens
the quality and reliability of the end product, reducing the number of defects present.
Automation also increases the overall effectiveness, avoiding the risk of human errors
and achieving repeatability. This is particularly useful to reduce the regression risk by finding
defects in the modified, but previously working, functionalities of the
system \cite{dustin1999automated}.

However, test automation does not always supersede manual testing. According to a number
of studies \cite{rafi2012benefits,wiklund2017impediments,taipale2011trade},
not all the testing tasks can be easily automated, such as those requiring extensive
knowledge in a specific domain, or they require a costly maintenance. In some cases, manual
testing can complement automation since, based on its unstructured nature, it could
potentially expose unexpected defects not considered in the previous stages within the
software life cycle.

%
%

\begin{figure*}[t]
\centerline{\includegraphics[scale=0.5]{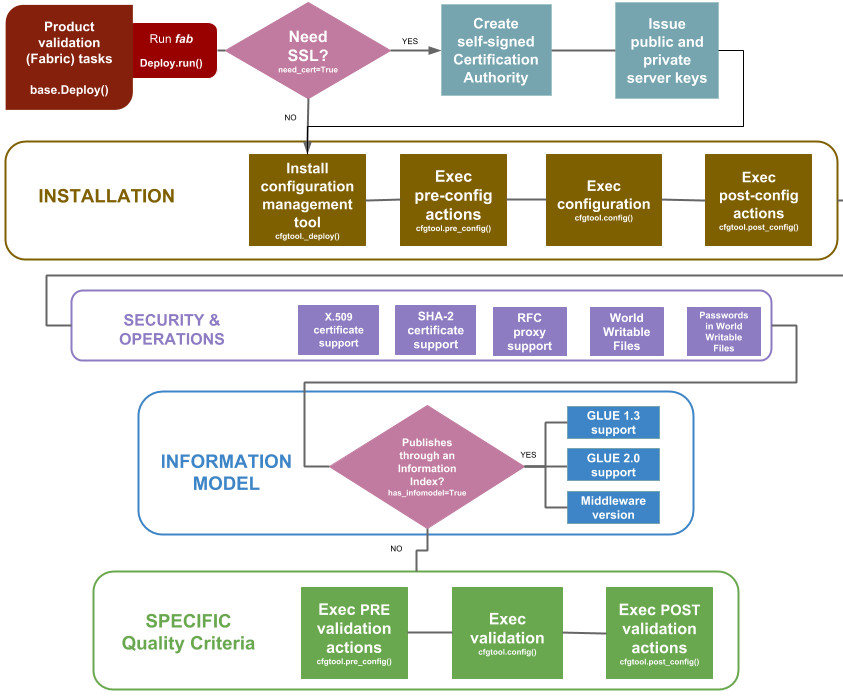}}
\caption{Product validation worflow in \texttt{umd-verification}.
\label{fig_umd_verification_flowchart}}
\end{figure*}

\section{\texttt{umd-verification}: an automated tool for the software validation process}
\label{section:implementation}


In order to automatize the software validation process within EGI, the essential component
would be a general purpose tool to manage the QC execution for each product validation.
This tool would execute the appropriate tasks for each requirement
analysis and, eventually, evaluate the obtained output values to judge whether the given
requirement has been fulfilled, allowing the process to stop depending on its criticality.


\subsection{Design considerations}

\subsubsection*{Infrastructure as Code deployments}
With the advent of Infrastructure as code (IaC) tools, the automated maintenance and
provision of services in an infrastructure is powered through a series of definition files,
which enforce the desired configuration of such services. Applying the IaC model to drive
the deployment part of the SWPP process would allow to have reliable, repeatable and
reusable configurations that supersede the traditional, less-efficient, manual guided
deployments. The solution to be implemented shall ensure the usage of common, well-known
IaC tools such as Ansible \cite{ansible} or Puppet \cite{puppet}.

\subsubsection*{Functional and integration testing}
The QC definition enforces the fulfillment of functionality testing, which covers the
newly added features and bugfixes. In this regard, the tool that orchestrates the 
QC validation shall be flexible enough to execute external scripts, wait for their
completion and approve the exit status.

Integration testing is needed whenever the product in validation interferes with additional
services while in operation. This type of testing requires more complex deployments, as the
related services must be in place in advance.

\subsubsection*{Dynamic provision of input parameters}
\begin{sloppypar}
Input parameters are needed in order to set up the diverse environments in which the
currently existing products are verified. The application managing the process needs to be
fed with several types of input parameter provision, such as run-time arguments, 
instantiation-time parameters and configuration files.
\end{sloppypar}


\subsubsection*{Inclusion of new products}
Based on the incremental trend of product adoption in the EGI software distributions, the
integration of new products into the automated solution proposed shall be an easy task. The
system shall provide a way to declare new products in a standard way, relying on an ubiquitous
language that requires little or no previous experience from the tester.

\subsection{Implementation of \texttt{umd-verification} tool}
\label{implementation:tool}

\texttt{umd-verification} tool \cite{umd-verification} is the solution proposed for the
automated, sequential validation of the requirements defined in the QC document. 
The tool is written in the Python programming
language \cite{python} and uses the Fabric library \cite{fabric} for a high-level management of the
system calls. \textit{Fabric-ed applications} are organized in tasks and have built-in
features such as remote executions and consistent argument passing via the command-line
\texttt{fab} tool.

\subsubsection*{Behind the scenes}

Figure \ref{fig_umd_verification_flowchart} shows the tool's workflow. Every new product
validation is represented by an instance of the customized Fabric base task, 
\texttt{base.Deploy}, which guides the process through four major execution blocks: i) 
installation and configuration, ii) security and operations, iii) information model, and iv)
specific QC. Note that, as already commented in Section \ref{sec:automation:qc}, 
documentation requirements need of human revision and thus are not being validated by the
application.


\begin{lstlisting}[language=python,basicstyle=\footnotesize\ttfamily, breaklines=true, frame=none,
caption={Python code snippet taken from the task validation of
\texttt{fts} product. The class attributes contain static information such as the relevant
pointers to enable the product's deployment using Puppet.}, captionpos=b, label=listing:fts, showstringspaces=false]
fts = base.Deploy(
    name="fts",
    doc="File Transfer Service (FTS) deployment.",
    need_cert=True,
    cfgtool=PuppetConfig(
        manifest="fts.pp",
        hiera_data=["fts.yaml", "fetchcrl.yaml"],
        module=[
            ("git://github.com/egi-qc/puppet-fts.git",
             "umd")]
    )
)
\end{lstlisting}

\begin{minipage}{\linewidth}
\begin{lstlisting}[language=python,basicstyle=\footnotesize\ttfamily, breaklines=true, frame=none,
caption={A complete task definition (\texttt{base.Deploy}) for the validation of
\texttt{cloud-info-provider} product (Python code). The task relies on an Ansible role for
the deployment, which needs a set of input variables that are defined within the
\texttt{pre\_config} method. The testing part is defined in an external configuration file
 (see Listing \ref{listing:qc_specific}), identified by the \texttt{cloud-info-provider} label.}, captionpos=b, label=listing:cloud-info-provider, showstringspaces=false]
from umd import base
from umd.base.configure.ansible import AnsibleConfig
from umd import config


class CloudInfoProviderDeploy(base.Deploy):
    def pre_config(self):
        # extra vars
        extra_vars = [
            "cloud_info_provider_os_username: demo ",
            "cloud_info_provider_os_password: secret ",
            "cloud_info_provider_os_release: %s "
            % config.CFG["openstack_release"],
            "cloud_info_provider_middleware: openstack ",
            "cloud_info_provider_conf_dir: /etc/cloud-info-provider ",
            "cloud_info_provider_bdii_dir: /var/lib/bdii/gip/provider"]
        self.cfgtool.extra_vars = extra_vars


cloud_info_provider = CloudInfoProviderDeploy(
    name = "cloud-info-provider",
    doc = "cloud-info-provider deployment using Ansible.",
    cfgtool = AnsibleConfig(
        role =  "https://github.com/egi-qc/ansible-role-cloud-info-provider",
        checkout = "umd",
        tags = ["untagged", "cmd"]),
    qc_specific_id = "cloud-info-provider")
\end{lstlisting}
\end{minipage}

As a result of being inherited from the base class \texttt{base.Deploy}, every product 
validation need to provide a set of class attributes that uniquely identify the product.
The code excerpt from Listing \ref{listing:fts} shows a sample implementation of a task
validation. One of these class attributes sets the next step in the workflow. A very
common requirement for Cloud and Grid services supported in EGI is to guarantee user
data protection by securing the connections using X.509 certificates \cite{housley1998internet}. 
Hence the definition of the \texttt{need\_cert} attribute, which when
enabled, issues a server certificate from a self-signed certification authority.

The first block, \textit{Installation}, addresses the deployment \textit{from scratch} of 
the product using an IaC solution. The \texttt{base.Deploy.\_deploy()} method first installs
the IaC tool and sets the required environment, such as generating parameter files and 
handling the module installation and its dependencies. The deployment is then triggered 
through the \texttt{base.Deploy.config()} method, with optional \textit{pre} and
\textit{post} steps that could have previously defined at instantiation time.

The \textit{Security and Operations} block is comprised of a set of basic security
assessments. This phase is specially significant for the secured products since it checks
the compliance with X.509 cryptographic standard and SHA-2 signatures \cite{sha2-nist}.

\begin{sloppypar}
Workload orchestration within EGI e-Infrastructure relies on the resource information
published by the providers. The \textit{Information Model} block ensures the presence of
published resource information, in GLUE format \cite{glue}, validated by the execution of an
external tool, \texttt{glue-validator} \cite{glue-validator}. As not all the supported products in 
UMD and CMD publish GLUE data, the class attribute \texttt{has\_infomodel} signals when
this requirement should be checked.
\end{sloppypar}

The last block, \textit{Specific Quality Criteria}, covers the functional and/or integration
testing of the product. Here, basic operation and new features and/or bugfixes included in
the release are tested. The class attribute \texttt{qc\_specific\_id} maps to the set of
checks, in the form of scripts, that must be executed. In the subsequent product validations,
these checks eliminate the regression risk as they are re-executed to ensure that the
previous working functionalities are kept.

Listing \ref{listing:cloud-info-provider} shows a more advanced usage of a validation task.
In this example the \texttt{base.Deploy.pre\_config()} method is overridden to set the
values of some parameters that need to be defined before the product's deployment using
Ansible. Moreover, the task is completed with a test definition through the
\texttt{qc\_specific\_id} attribute.

\subsubsection*{Support for new software components}
\label{implementation:support}

\begin{sloppypar}
One of the key design considerations of \texttt{umd-verification} application was to
ease the addition of new product validations, while relying on a powerful and ubiquitous
language. As described in the section above, the usage of the Python language matched both
design requirements with the only caveat of assuming certain degree of experience in
Python programming, specially in the case of very customized and complex configurations.
\end{sloppypar}

The simplest case would directly inherit from the base class \texttt{base.Deploy}, while
more complex scenarios would create a child class, overriding the necessary class
methods and attributes as shown in Listing \ref{listing:cloud-info-provider}.
In either case, a new task definition is added by filling in the mandatory attributes, 
consisting in the \texttt{name} and \texttt{description}, the IaC configuration represented
by the \texttt{cfgtool} attribute and the associated test checklist identified by the
\texttt{qc\_specific\_id} attribute. Enabling or disabling the optional attributes
\texttt{need\_cert} and \texttt{has\_infomodel} further define the validation task and,
consequently, the workflow to be followed in the \textit{Security} and
\textit{Information Model} blocks.

\begin{sloppypar}
Deployment settings vary with respect to the IaC tool in use, having each a different 
object class that takes over the deployment based on the parameters passed, and 
accessible through the \texttt{base.Deploy.cfgtool} attribute. Listings
\ref{listing:fts} and \ref{listing:cloud-info-provider} use different \texttt{cfgtool}
objects, representing Puppet and Ansible respectively.
\end{sloppypar}

Product testing needs a definition where the checklist of tests are listed in
order to be triggered in the task validation. Listing \ref{listing:qc_specific} shows
an excerpt of the configuration file used for the test definitions. Tests are
categorized by the QC requirement --either \texttt{QC\_FUNC\_1} or
\texttt{QC\_FUNC\_2}-- and defined by the test description, location and arguments.

\begin{lstlisting}[language=yaml, basicstyle=\footnotesize\ttfamily, breaklines=true,
caption=\texttt{cloud-info-provider} YAML test definition., label=listing:qc_specific, captionpos=b]
cloud-info-provider:
    qc_func_1:
        - test: "bin/bdii/client-test.sh"
          description: "GLUE2 ldapsearch check."
          args: "ldapsearch-site-bdii-cloud"
\end{lstlisting}



\begin{figure*}[t]
\centerline{\includegraphics[scale=0.4]{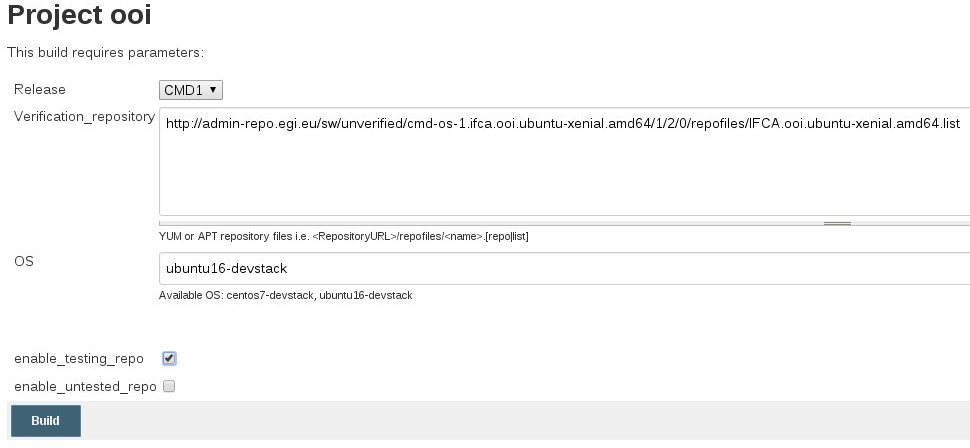}}
\caption{Product validation form in Jenkins. The input (runtime) parameters are
comprised of the software release --from UMD or CMD distributions--, the operating system 
and the URL of the verification repository that contains the software packages to be
validated.
\label{fig_jenkins_form}}
\end{figure*}

Based on the above guidelines, new product adoption within the 
\texttt{umd-verification} tool is \textbf{not a costly task whenever the IaC modules
and tests are already available}, either provided by the TPs or individuals that share
their work publicly. Both deployment and testing are time-consuming tasks, if
performed from scratch, that require a great deal of expertise in the candidate
product and, additionally, in the IaC tool being used.

\section{Evidence of the \texttt{umd-verification} adoption}
\label{section:evidence}

\subsection{Continuous Integration implementation}

\texttt{umd-verification} is suitable for being integrated in a Continuous Integration
(CI) pipeline. The CI system fires up the virtual resource, sets up the application, 
triggers the execution with the appropriate runtime parameters and, finally, tears down
the provisioned resource. All these steps are condensed in a \textit{job} definition
within the Jenkins CI service \cite{jenkins-egi} for each product in the catalogue. 
Figure \ref{fig_jenkins_form} shows a sample form in Jenkins CI that, on submission,
will trigger the validation process leveraging the \texttt{umd-verification} tool. The
run-time parameters passed are commonly the ones showcased in the figure, consisting in
the EGI distribution and Linux OS, and the additional repositories, such as the one
containing the candidate version of the software product.

The usage of a CI service to take over the validation of products, notably
\textbf{hides the inner complexity of the validation process} --resource provisioning,
\texttt{umd-verification} deployment and execution--, \textbf{allowing a non-expert
usage}. 

\subsection{Time efficiency for the validation process}

\begin{figure}[t]
\centerline{\includegraphics[scale=0.36]{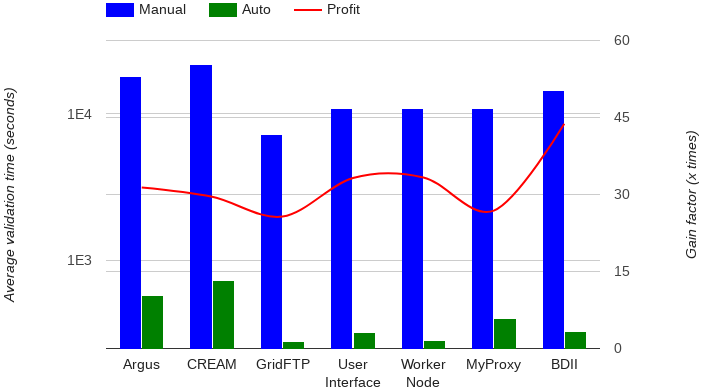}}
\caption{Automated vs Manual validation process times. Time values on the vertical 
axis use a logarithmic scale to better showcase the important differences of time
completion for both types of validation processes. Manual values are not as accurately
estimated as the automated ones. Whilst the latter have been obtained from the CI
service, the manual values were extracted from the validation reports uploaded to the
EGI Document Database \cite{egi-docdb}.}
\label{fig_auto_vs_manual}
\end{figure}

The paramount benefit of automating the validation process via the 
\texttt{umd-verification} application is the time efficiency. Combined with the 
automated resource provisioning, provided by the CI implementation previously described,
this efficiency raises even higher.

As it was mentioned in the statement of the problem in Section \ref{sec:automation:problem},
back in the days of the manual validation process \cite{Mario2014}, a common completion time
was estimated to be 1 or 2 days. With the new approach the validation process takes a few
minutes, although this duration is tightly related to the deployment requirements of each
software component, as some products need additional services for the testing phase. Therefore,
the time required to add support for a new product within the \texttt{umd-verification} tool
may be costly whenever there is no availability of IaC modules. Otherwise, as it was shown in
Section \ref{implementation:support}, the definition of the new product in Fabric is an 
immediate task. 

The data displayed in Figure \ref{fig_auto_vs_manual} compares the validation time of both
approaches for a set of UMD products, showcasing the profit percentage obtained with the 
automated process. The results show an \textbf{average factor of 32 in
the time efficiency of the validation process} with the adoption of the automation process
described throughout this paper. 

\subsection{IaC knowledge base}

\begin{figure}[t]
\centerline{\includegraphics[scale=0.5]{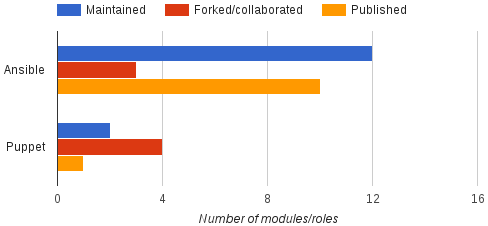}}
\caption{Ansible roles and Puppet modules being maintained, forked and published in the
official repositories by the EGI validation team.
\label{fig_iac_modules}}
\end{figure}

\begin{sloppypar}
One of the requirements imposed when supporting a new product validation in the
\texttt{umd-verification} application is the usage of an IaC solution for its deployment.
Since the adoption of automation, the EGI validation team maintains a public
repository \cite{egi-qc} with a collection of Ansible and Puppet modules resultant from the
validation process. Figure \ref{fig_iac_modules} provides an overview of the work being done
in this regard and referred as maintained --modules created and supported by the EGI
validation team--, forked --modules modified and contributed to upstream-- and
published --modules contributed to the official Ansible \cite{egi-qc-ansible} and 
Puppet \cite{egi-qc-puppet} community repositories--.
\end{sloppypar}

As self-documentation code, \textbf{IaC modules shaped in the validation phase can be then
re-used in a reproducible way in future deployments}. As a result, within the EGI
e-Infrastructure, resource center operators can make use of those modules to deploy the
products in the EGI catalogue. This contrasts with the previous procedure, where deployments
done in the validation phase could not be easily reproduced: they were locally addressed
by the tester, with the only reference of a set of non-structured annotations being included
in the verification report.

\subsection{Release Candidate validation}

\begin{figure*}[t]
\centerline{\includegraphics[scale=0.45]{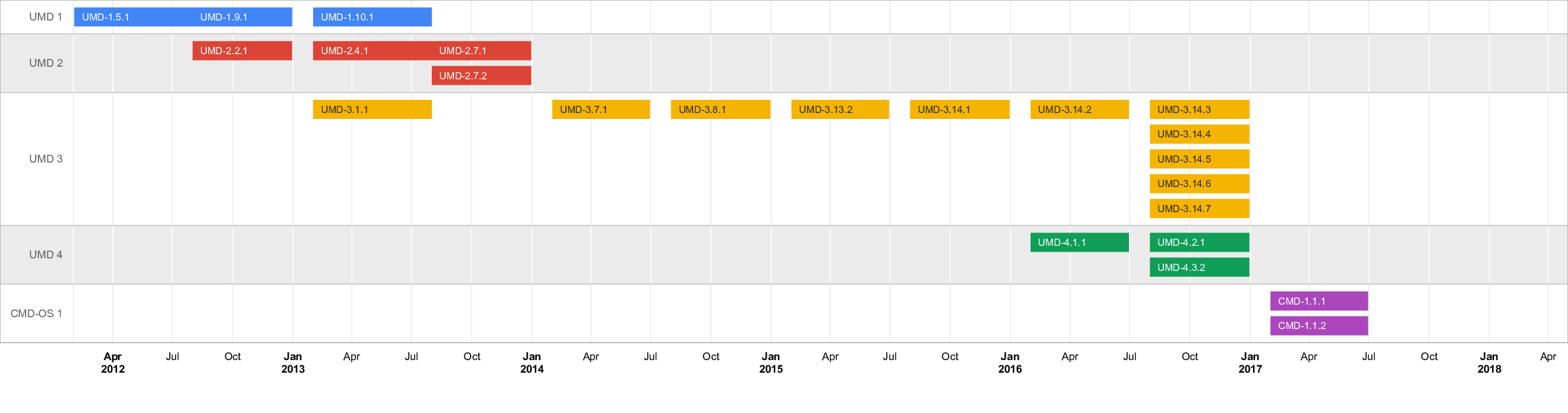}}
\caption{UMD and CMD revision releases that fixed previous releases with package dependency
issues included. As of April 1st 2018, no revision release was needed to solve package
unmet dependencies.
\label{fig_release_candidate}}
\end{figure*}

Contiguous software validations are packed in releases, each defined by a version number
that reflects its purpose either as revision, minor or major release. Every release is checked by
the EGI validation team before being announced as production-ready, following the
\textit{Release Candidate} (RC) procedure.

\begin{sloppypar}
The early RC implementation relied on a script file that verified the installation of the EGI 
product catalogue \cite{rc-tester}. The list of products was manually maintained, adding or
removing entries in the script as the catalogue evolved. With the advent of the 
\texttt{umd-verification} tool and the adoption of IaC capabilities, the validation of the
RCs is eventually tackled using an Ansible role \cite{rc-ansible-role}. This new implementation
\textbf{fetches dynamically the whole set of packages of either UMD or CMD repositories,
to detect any unresolved dependency} that might be introduced by the new
packages that take part of the release. Whenever detected, the validation team fixes the
dependency issue and re-runs \texttt{umd-verification} tool until all the packages in
the repository are properly installed. 
\end{sloppypar}

\begin{sloppypar}
The dynamic gathering of packages profits from Linux package management utilities, thus
there is no need to maintain a static list of software packages to install for each RC. 
Furthermore, in the past this list did not contain the complete set of packages but only the
ones that refer to the main products in the catalogue. As a result, there was a potential
risk of uninstallable packages living at the EGI repositories. As it can be seen in Figure 
\ref{fig_release_candidate}, almost 25 revisions --since 2012-- were explicitly devoted to
solve troubles in package dependencies. With the new implementation --back in July 2017--
the number of revision releases meant for dependency resolution dropped to zero. 
\end{sloppypar}



\section{Conclusions}
\label{conclusions}

Based on the growing demands of adopting new products and supporting existing ones in the
UMD and CMD releases, the validation of the conformance criteria has to move forward to an
automated process. The suitability of the requirements currently existing in the EGI QC
document to be addressed programmatically paved the path to the  implementation and further
integration of the \texttt{umd-verification} tool within the EGI SWPP. The current set of
existing products in UMD and CMD repositories are being progressively integrated in the 
new automated process, often at the cost of developing the required automated deployment
and test cases whenever they are not provided directly by the technology provider or 
shared within the community. However, for those products already integrated, the evidence
of improvement has been demonstrated both in terms of efficiency, as the process validation
times are clearly optimized, and effectiveness, by getting rid of the likelihood of human
error.

Completion time efficiency is the most apparent benefit of adopting automation, shortening
the process in an average factor of approximately 32 when compared with the reported time of
traditional manual validations. This implies less human effort than the former approach, now
reallocated to integration and maintenance activities, being in a better position to confront 
unexpected demands of product validations. Growing needs of manpower are no longer the  
solution to high demands as, once the cost of integrating the product in the
\texttt{umd-verification} tool is assumed, it will require little or no human intervention.

The programmatic evaluation of the EGI QC requirements combined with the adoption of IaC 
solutions achieved repeatability and reproducibility in the process of validating software.
In particular, IaC modules make the deployment of the products to be reproduced and shared,
contributing to the creation of a knowledge base within the community. Just as the IaC
modules facilitate the products' deployment to non-experienced users,
so does the \texttt{umd-verification} solution with the validation of conformance criteria.
Testers not familiarized with a given product can now take over its validation
without any expert intervention and, as a result of enabling \texttt{umd-verification} 
within a CI scenario, even testers with no previous experience with the tool ought to
complete the validation process. This accessibility has remarkably reduced the risk of
expert dependence, which was tightly associated to the former manual process.

\begin{acknowledgements}
\begin{sloppypar}
This work has been partially funded by the EGI-Engage project (Engaging the Research
Community towards an Open Science Commons) under grant agreement No. 654182. The authors 
are especially grateful to EGI.eu's colleagues Enol Fern\'andez, for his contributions to
the \texttt{umd-verification} codebase, and Vincenzo Spinoso, for his support in the 
tool integration within the EGI Software Provisioning Process.
\end{sloppypar}
\end{acknowledgements}

\bibliographystyle{unsrt}
\bibliography{main}


\end{document}